# Evolutionary game on any hypergraph

Dini Wang[1,2], Peng Yi[1,2,*], Yiguang Hong[1,2], Jie Chen[1,2], Gang Yan[2,3,*]

[1] *College of Electronic and Information Engineering, Tongji University, Shanghai, 201804, P. R. China*

[2] *National Key Laboratory of Autonomous Intelligent Unmanned Systems, MOE Frontiers Science Center for Intelligent Autonomous Systems, and Shanghai Research Institute for Intelligent Autonomous Systems, Tongji University, Shanghai, 201210, P. R. China*

[3] *School of Physical Science and Engineering, Tongji University, Shanghai, 200092, P. R. China*

[*] Correspondence: yipeng@tongji.edu.cn; gyan@tongji.edu.cn

## Abstract

Cooperation plays a fundamental role in societal and biological domains, and the population structure profoundly shapes the dynamics of evolution. Practically, individuals behave either altruistically or egoistically in multiple groups, such as relatives, friends and colleagues, and feedbacks from these groupwise interactions will contribute to one's cognition and behavior. Due to the intricacy within and between groups, exploration of evolutionary dynamics over hypergraphs is relatively limited to date. To uncover this conundrum, we develop a higher-order random walk framework for five distinct updating rules, thus establishing explicit conditions for cooperation emergence on hypergraphs, and finding the overlaps between groups tend to foster cooperative behaviors. Our systematic analysis quantifies how the order and hyperdegree govern evolutionary outcomes. We also discover that whenever following a group-wisdom update protocol, choosing a high-fitness group to interact equally within its members, cooperators will significantly prevail throughout the community. These findings underscore a crucial role of higher-order interaction and interdisciplinary collaboration throughout a broad range of living systems, favoring social prosperity.

## Introduction

Cooperative behaviors manifest across various scales, spanning from individual cells to expansive social communities[1-6]. Despite this, natural selection typically favors competition, posing a challenge to cooperation unless a robust mechanism intervenes[7-10]. Network reciprocity recognizes that certain population structures can facilitate the prevalence of cooperators, even without direct incentives[8,11-15]. While extensive research has explored pairwise interactions through simulations, approximations, and analytical solutions[16-26], it is crucial to acknowledge that spatial interactions or social connections often exhibit multi-group (higher-order) structures[27-32] that can influence collective cooperation as shown in recent few work[33-36]. However, the potential of these higher-order structures to offer innovative mechanisms for cooperation and their influence on the emergence of cooperation remain not fully understood. Here, we systematically explore the evolutionary dynamics of public goods games (PGGs)[14,37,38] on any hypergraphs and offer the analytical conditions for the emergence of cooperation with higher-order interactions.

We use a hypergraph[27-32], a generalization of a simple graph, to model the higher-order relationship. In a simple graph an edge connects exactly two vertices, whereas in a hypergraph an edge, called hyperedge hereafter, can join more than two vertices (see Fig. 1. (A)). The order of a hyperedge $e$, denoted by $g_e$, is the number of vertices in the hyperedge. The hyperdegree $k_i$ of vertex $i$ is the sum of the weights of adjacent hyperedges around $i$. Each hyperedge $e$ is given a weight $w_e$ to capture the importance of the group. For example, in Fig. 1a, the hyperedge $\beta$ contains four vertices hence $g_\beta = 4$; vertex 2 is involved in the hyperedge $\alpha$ with weight $w_\alpha = 2$ and in $\beta$ with weight $w_\beta = 1$, hence its hyperdegree $k_2 = 3$.

Evolution of cooperation on hypergraphs consist of two key steps during each game round. First, the vertices in the same hyperedges play the PGG; Second, a randomly-selected vertex needs to decide whether it changes its strategy (from cooperator to defector, or vice versa) based on its own fitness and neighbors' information. Regarding the first step, within each hyperedge featuring $g$ vertices, each cooperator invests a fixed cost $c$ into the public goods, whereas defectors make no contribution. Subsequently, all participants receive an equal share of the benefits $b$, calculated by multiplying the total investment by the enhancement factor $r = b/c$. Alternatively, we can interpret the cooperator as paying $c/g$ for each participant, ensuring that each beneficiary gain a profit of $b/g$ (see Fig. 1. (B)). The payoff from each game needs to be averaged by the weights of the hyperedges to derive the final payoff $f_i$ for individual $i$ (see Fig. 1. (C)). The fitness is then expressed as $F_i = 1 + \delta f_i$, where $\delta$, a positive parameter approaching zero, signifies weak selection.

Regarding the second step, *i.e.,* strategy update, we explore two categories of rules governing the evolutionary games on hypergraphs. The first category extends existing death-birth, imitation, and pair-comparison rules to higher-order implementations, denoted as higher-order death-birth (HDB), higher-order imitation (HIM), and higher-order pair-comparison (HPC). The second category is new and specifically designed for higher-order structures, highlighting the group dynamics through group inner comparison (GIC) and group mutual comparison (GMC).

Our primary objective is to determine to what degree cooperation is advantageous in a specific hypergraph. In the absence of mutation, the evolutionary system ultimately reaches an absorbing state where either all individuals cooperate or all individuals defect after a sufficient number of game rounds. The fixation probability of cooperation, denoted as $\rho_{\mathrm{C}}$, represents the likelihood that a randomly arising cooperator can transition the entire population from defection to cooperation. In the case of neutral drift ($\delta = 0$), the fixation probability equals $1/N$, where $N$ is the number of vertices in the system. Therefore, if $\rho_{\mathrm{C}} > 1/N$, the evolutionary mechanism favors the emergence of cooperation. Consequently, our investigation is centered on determining the critical enhancement factor for cooperation—specifically, the smallest value of $r$ that makes $\rho_{\mathrm{C}}$ exceed the baseline of $1/N$.

## Results

### Random walks, PGGs and evolution on hypergraphs

Higher-order random walk offers a unique perspective for understanding the evolutionary game on hypergraphs. Consider vertex 2 in Figure 1 as an example. This vertex engages in two PGGs within hyperedges $\alpha$ or $\beta$, respectively (Fig. 1. (B)), obtaining the payoff as the weighted average of its adjacent vertices' contributions over the hyperedges' weights (Fig. 1. (C)). After all vertices complete a round of the game, we extract the payoff coefficient to construct the *interaction* graph as shown in Fig. 1. (D). This process is related to random walks with self-loops on the hypergraph. Indeed, a walker starts from vertex 2 and arrives in hyperedge $\alpha$ or $\beta$ with a probability proportional to their weights (Fig. 1. (E)), then evenly traverses each member within the hyperedge (Fig. 1. (F)). The one-hop transition probability from vertex 2 to its neighbors is exhibited in Fig. 1. (G); thus, we obtain, from the viewpoint of random walk, the *weighted* effective graph $\hat{G}$ (Fig. 1. (H)). We surprisingly find that $\hat{G}$ is identical with the interaction graph. For random walks with no self-loops, we construct another effective graph $\tilde{G}$ (Fig. 1. (I-L)). Detailed constructions of $\tilde{G}$ and $\hat{G}$ are shown in Supplementary Materials Section 1.

For analysis of evolutionary dynamics on hypergraphs, we introduce a novel $(n, m)$-random walk pattern to represent a path of $n$ hops on graph $\tilde{G}$ followed by $m$ hops on graph $\hat{G}$, and the probability of such $(n, m)$-walks from vertex $i$ to vertex $j$ is denoted by $p_{ij}^{(n,m)}$. Thus, the hyperdegree-weighted probability of an $(n, m)$-walk starting from a vertex and returning itself can be described by $\theta^{(n,m)} = \sum_{i\in\mathcal{V}} \pi_i p_{ii}^{(n,m)}$ where $\mathcal{V}$ is the set of all vertices in the hypergraph, and $\pi_i = k_i / \sum_{j\in\mathcal{V}} k_j$ is the normalized hyperdegree of vertex $i$. The parameter $\theta^{(n,m)}$ represents the recurrence strength of random walks on the hypergraph.

Furthermore, the constructed graph also provides valuable insights into the process of strategy updating, as illustrated in Table 1. In HDB updating, following the game interaction, a vertex is randomly selected for replacement. An adjacent hyperedge is then chosen with probability proportional to fitness times hyperedge weight, and subsequently, a vertex in this hyperedge is selected with a probability proportional to the vertex's fitness. The chosen vertex reproduces into the vacant position. From the perspective of random walk, the influence of vertex $j$ on vertex $i$ can be quantified by the probability of a random walk from $i$ to $j$ under neutral drift (*i.e.*, $\delta = 0$). As such, the replacement graph becomes $\tilde{G}$ (without self-loops) under HDB. HIM is akin to HDB, with the distinction that in HIM, the to-be-replaced vertex itself is also considered in the selection process, whereas in HDB, it is not. This results in an equivalent replacement event on the graph $\hat{G}$ (with self-loops). In HPC updating, an adjacent hyperedge is selected based solely on the hyperedge's weight, and a vertex in this hyperedge is chosen uniformly at random. The to-be-replaced vertex then compares its fitness to that of the selected vertex to decide whether to adopt its strategy. And the similar details of GIC and GMC updating rules are depicted in Table 1. From the aspect of equivalent random walk, under GIC rule, the strategy updates equivalently occur on the graph $\hat{G}$ (with self-loops), whereas under HPC and GMC rules, the replacement graph becomes $\tilde{G}$ (without self-loops). In essence, the five updating rules exhibit varying degrees of greed during the decision-making process for choosing an adjacent hyperedge and selecting a vertex within that hyperedge.

**Analytical conditions of cooperation emergence on hypergraphs**

Using an individual-based mean-filed approach, we analytically obtain the critical thresholds of enhancement factor for PGGs on any hypergraph and for all five updating rules (see Supplementary Materials Sections 3 and 4 for the mathematical derivations). As shown in Fig. 2. (A), for such complicated evolutionary process of PGGs on hypergraph, we find, surprisingly, that the critical thresholds $r^*$ can be captured by two factors: one is the property of $(n, m)$-random walks, $\theta^{(1,1)}$ or $\theta^{(0,2)}$; and the other is the hypergraph's topological properties, hyperedges' average order $\langle g\rangle$ or hyperdegree heterogeneity $\tilde{N} = N\langle k\rangle^2/\langle k^2\rangle$ (*i.e.*, smaller $\tilde{N}$ indicates higher heterogeneity).

We validate our analytical results with Monte Carlo simulations on hypergraphs whose orders and hyperdegrees are either homogeneous or heterogeneous. As shown in Fig. 2. (B-D), the critical enhancement factors for updating rules HPC and GIC are the same and larger than $\langle g \rangle$. Note that there is no social dilemma in a PGG with $g$ players if $r > g$, because each player gets benefit larger than investment. Thus, we also derive critical enhancement factor normalized by hyperedge order in Supplementary Materials Section 6. The result indicates that HPC and GIC rules are detrimental to the cooperation on hypergraphs. In contrast, the critical enhancement factor for GMC is significantly smaller than that of other rules, implying that selecting a high-fitness group and interacting equally with the members in the group are helpful for the emergence of collective cooperation.

Furthermore, taking the partial derivative of the critical enhancement factor with respect to hypergraph's key properties, we obtain the mathematical expressions capturing the influence of hyperedges' order and hyperdegree on cooperation (see Supplementary Materials Sections 5). As shown in Fig. 2. (E-G), hypergraphs with either smaller order or smaller hyperdegree are easier to produce collective cooperative behaviors; the heterogeneity of hyperdegrees can promote cooperation yet only under GMC rule.

For analysis of multi-group dynamics in human society, we also investigate four empirical hypergraphs, ranging from scientific coauthorships among research scholars in DBLP and MAG, to online relationships among members in the forums of Mathematics Stack Exchange and Ask Ubuntu. The topological properties of these real-world hypergraphs are illustrated in Table 1, involving weight, population size, count of hyperedges as well as averages of hyperdegree and order. We analyze the prospects of cooperative behaviors under five update protocols, when individuals engage in the PGG within each group, by critical enhancement factors derived from Fig.2. (A). To verify the accuracy of the critical values, we simulate the evolutionary process until the system is fixed. As shown in the last column of Table 1, the fixation probability times the population size all approaches one, implying a perfect prediction of the real enhancement factor required for cooperation. This is due to, disregarding game effects, the probability that drives the population toward homogeneity is just $1/N$.

**Overlaps between hyperedges promote cooperation**

Overlap between any two hyperedges is a defining property of hypergraphs. Here, we introduce $C_{\mathrm{ol}}$ to quantify the overlap property, referring to the number of one-hop walks in the overlapping sets. To analytically explore the influence of overlaps, we consider regular hypergraphs (*i.e.*, all vertices have the same hyperdegree and all hyperedges have the same order) whose $C_{\mathrm{ol}}$ is described in Methods

Section III. This allows us to recast the mathematical expressions of the critical enhancement factors to involve $C_{\mathrm{ol}}$ (see Supplementary Information Section 5), which indicate that the overlaps between hyperedges can significantly promote cooperation. We numerically validate the theoretical predictions and demonstrate that the critical enhancement factor indeed decreases when overlapping size increases (Figure 3a-d). From the perspective of random walk, higher overlaps among hyperedges provide more pathways for the two-hop walks from one vertex and back to itself, increasing the probabilities of $\theta^{(1,1)}$ and $\theta^{(0,2)}$ thus lowering the critical thresholds $r^*$.

**Comparison between hypergraphs and simple graphs**

To further demonstrate the significant role of higher-order interaction, we next compare PGGs on a hypergraph and on its *trivially* reduced simple graph (*i.e.*, preserving only the pairwise connection relationships, as shown in Fig. 4. (F)). In PGGs on a simple graph, each vertex plays a game with its nearest neighbors, hence a vertex of degree $d$ participates $d+1$ games in each round[14]. We derive the theoretical values of critical enhancement factors for PGGs on simple graphs for updating rules DB, IM and PC (see Supplementary Information Section 7), and show the numerical calculations of critical thresholds for four real hypergraphs (see Data Availability) and their trivial simple graphs. We observe that PGGs on simple graphs exhibit significantly higher critical thresholds than their hypergraph counterparts (Fig. 3. (H)), and the patterns of threshold ranking are also significantly different (Fig. 4). This suggests that higher-order interactions typically promote the dissemination of cooperative behaviors.

## Conclusion

In summary, this work expanded the mathematical framework for studying the evolutionary dynamics of public good games on simple graphs to hypergraphs. We derived analytical conditions for the emergence of cooperation and unveiled the impact of higher-order interactions on the dissemination of cooperative behaviors. Our findings indicate that the heterogeneity of hyperdegrees promotes cooperation under specific rules of strategy update, and the overlaps between hyperedges generally foster cooperation. This paradigm shift from pairwise to higher-order structures deepens our understanding of evolutionary dynamics in complex social and biological systems.

## Methods

The model and mathematical methods are summarized in the following and the complete derivations are provided in Supplementary Information.

**Public goods game on hypergraph**

Population structure is represented by a hypergraph where each hyperedge has a weight and can contain more than two vertices. Public goods game (PGG) occurs within each hyperedge, and in each round each vertex decides whether to be a cooperator or a defector. Each cooperator invests the amount $c$ into a public pot and the defectors invest zero. The total investment in this pot is multiplied by an enhancement factor $r$ then evenly divided among the vertices in the same hyperedge. The enhancement factor $r$ is also called benefit-to-cost ratio and $b = c \cdot r$ represents benefit.

We show above that, to capture the evolutionary dynamics of PGG, a hypergraph can be projected to two weighted simple graphs, $\tilde{G}$ without self-loops and $\hat{G}$ with self-loops, from the perspective of random walk. The transition probabilities of a random walk from vertex $i$ to vertex $j$ are $\tilde{p}_{ij}$ and $\hat{p}_{ij}$ respectively. We also introduce the new $(n, m)$-hop random walk, representing a walk of $n$ hops in graph $\tilde{G}$ followed by $m$ hops in graph $\hat{G}$. The probability of such $(n, m)$-hop random walks from vertex $i$ to vertex $j$ is denoted by $p_{ij}^{(n,m)}$.

The system state is given by the vector $\boldsymbol{s} = (s_1, s_2, \ldots, s_{N-1}, s_N)$, where $s_i$ represents the probability of vertex $i$ being a cooperator. Hence, $0 \le s_i \le 1$, and $s_i = 1$ for cooperator while $s_i = 0$ for defector. Without loss of generality, we set the cost $c = 1$ and focus on the enhancement factor $r$. Hence, the benefit of any vertex $i$ is the summation of $s_j r \hat{p}_{ij}$ over all of its neighors $j$, *i.e.*, the payoff of vertex $i$ is quantified by

$$f_i(\boldsymbol{s}) = \sum_{j \in V} \hat{p}_{ij} s_j \, r - s_i. \tag{1}$$

The fitness of $i$ is $F_i = 1 + \delta f_i$, where $\delta \ll 1$ represents the intensity of selection.

**Derivation of critical enhancement factor**

Here we briefly show the derivation for HDB updating, and the complete derivations for other four updating rules are described in Supplementary Information Section 4. By neglecting the dynamical correlations between the states of the neighbors, the individual-based mean-field approach yields the fixation probability of cooperation for HDB as

$$\rho_{\mathrm{C}} = \frac{1}{N} + \frac{\delta}{N}\int_0^\infty \Bigg[-\sum_{i,u\in\mathcal{V}} \pi_i s_i^u(t)\big(s_i^u(t)^{(0,0)} - s_i^u(t)^{(2,0)}\big) + r \cdot \sum_{i,u\in\mathcal{V}} \pi_i s_i^u(t)\big(s_i^u(t)^{(0,1)} - s_i^u(t)^{(2,1)}\big)\Bigg]_{\delta=0} dt + O(\delta^2)\,,$$

where $s_i^u(t)^{(n,m)} = \sum_{j\in\mathcal{V}} p_{ij}^{(n,m)} s_j^u(t)$ represents the expected state of $(n,m)$-hop neighbors of vertex $i$ at time $t$ when vertex $u$ is the initial single cooperator, and $\pi_i = k_i/\sum_{i\in\mathcal{V}} k_i$ represents the normalized hyperdegree of vertex $i$. Under neutral drift, $\delta = 0$, the state of vertex $i$ at time $t$ can be described as $[s_i^u(t)]_{\delta=0} = \sum_{j\in\mathcal{V}} \tilde{\xi}_{ij}(t)\, s_j^u(0) = \tilde{\xi}_{iu}(t) s_u^u(0)$ where $\tilde{\xi}_{ij}(t)$ is the probability of strategy alteration during the time interval. Taking into account integral transform and elimination, we have

$$\rho_{\mathrm{C}} = \frac{1}{N} + \frac{\delta}{2}\Bigg[-\Bigg(\sum_{u\in\mathcal{V}} \pi_u p_{uu}^{(0,0)} + \sum_{u\in\mathcal{V}} \pi_u p_{uu}^{(1,0)} - 2\sum_{u\in\mathcal{V}} \pi_u^2\Bigg) + r \cdot \Bigg(\sum_{u\in\mathcal{V}} \pi_u p_{uu}^{(0,1)} + \sum_{u\in\mathcal{V}} \pi_u p_{uu}^{(1,1)} - 2\sum_{u\in\mathcal{V}} \pi_u^2\Bigg)\Bigg] + O(\delta^2).$$

The critical threshold is the smallest $r$ to make $\rho_{\mathrm{C}} > 1/N$, hence we obtain the critical enhancement factor $r^*$ as shown in Fig. 2. (A) by setting the term in the square bracket equal zero. The derivation details are described in Supplementary Information Section 3.

**Quantifying the overlaps between hyperedges**

To investigate the impact of overlaps between hyperedges on the emergence of cooperation, we concentrate on homogeneous hypergraphs, where all vertices share the same hyperdegree $k$, and all hyperedges possess the same order $g$. The overlap can be quantified by the expression:

$$C_{\mathrm{ol}} = \frac{\sum_{i,j\in\mathcal{V}} \sum_{\alpha,\beta\in\mathcal{E},\beta\neq\alpha} b(i,\alpha)b(j,\alpha)b(j,\beta)b(i,\beta)}{Nk(k-1)g}$$

where $b(i,\alpha) = 1$ if $i \in \alpha$ and 0 otherwise $\mathcal{E}$ represents the hyperedge set. The numerator calculates the count of pairs of hyperedges overlapping two vertices $i$ and $j$ (allowing $i = j$), and the denominator is the normalization factor. When each pair of adjacent hyperedges overlaps over only one vertex, the hypergraph has the minimum overlapping strength $C_{\mathrm{ol}} = 1/g$. In contrast, $C_{\mathrm{ol}}$ reaches the maximum value of 1 when the overlapping size equals the hyperedges' order.

## Code availability

Code that supports the findings of this study is available in *GitHub* via the link

https://github.com/diniwang/EvolutionaryGame-Hypergraph.

## Data availability

Several empirical hypergraphs are used in this study. A congress bill dataset, where a vertex is a congressperson and a hyperedge is comprised of the sponsor and co-sponsors of a bill. Two human contact datasets, in a primary school and a high school, where each vertex is a student and each hyperedge is a group of students in close proximity during an interval. Two email networks from Enron company and a European institution, where a vertex is an email account and a hyperedge involves the sender and all recipients of an email. Two coauthorship networks, DBLP and MAG, where each vertex is an author and each hyperedge represents a publication with multiple coauthors. Two online interaction networks, in the forums Mathematics Stack Exchange and Ask Ubuntu respectively, where each vertex is a user and each hyperedge represents a thread involving multiple users. The empirical hypergraph data can be accessed via https://github.com/arbenson/ScHoLP-Data.

## Author Contributions

G.Y. and P.Y. conceived the project; D.W. performed mathematical analysis and numerical simulations with help from P.Y. and G.Y.; G.Y., P.Y., Y.H. and J.C. analyzed the results; D.W., G.Y. and P.Y. wrote the paper with input from Y.H. and J.C.. All authors contributed to all aspects of the project.

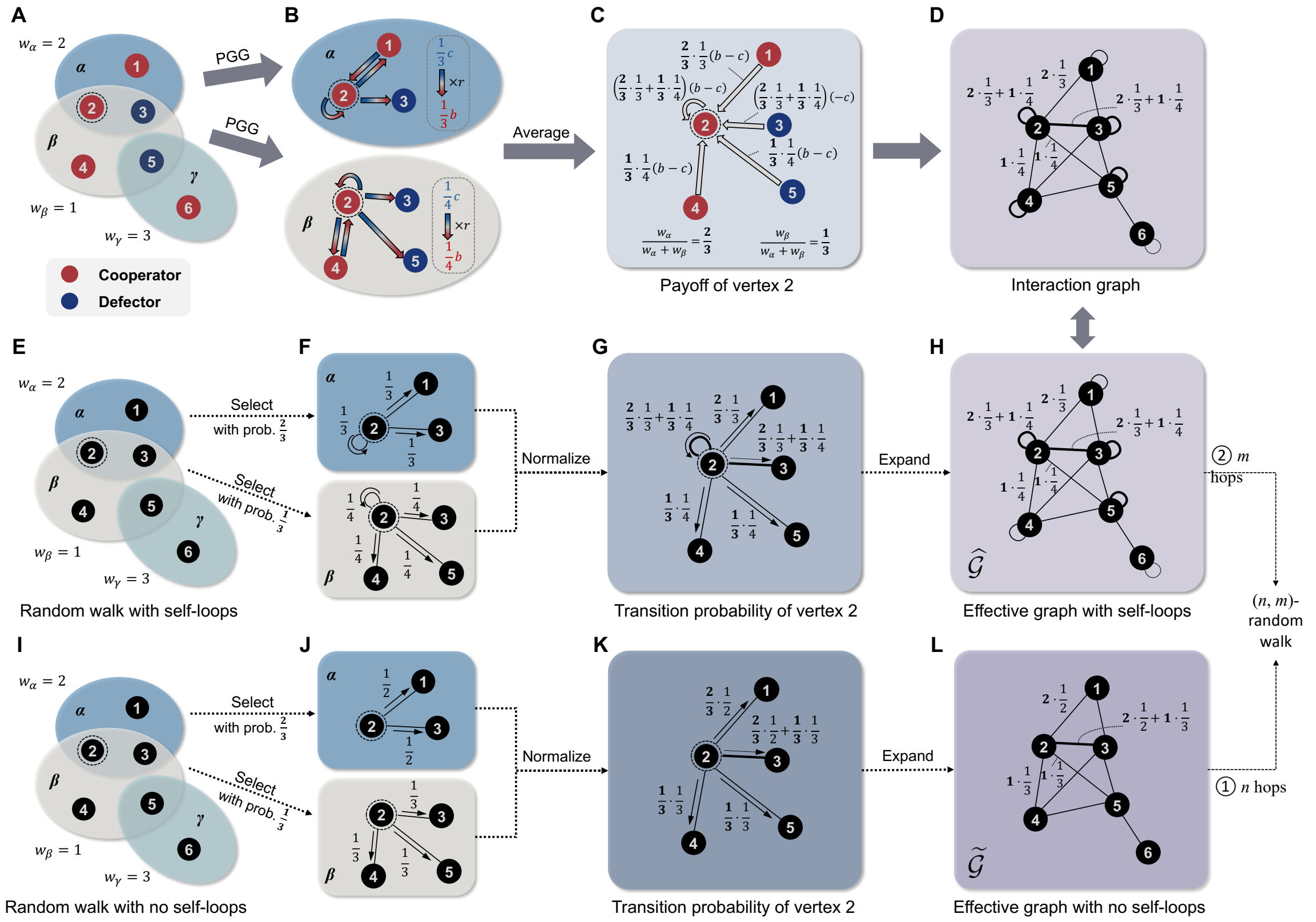

**Fig. 1. Public goods game (PGG) on hypergraph and random walk based effective graphs. A,** A hypergraph consisting of six vertices and three hyperedges ($\alpha$: {1,2,3}, $\beta$: {2,3,4,5}, and $\gamma$: {5,6}) whose weights are $w_\alpha$, $w_\beta$, and $w_\gamma$ respectively. Each vertex can be a cooperator (red) or a defector (blue). **B,** Vertex 2 belongs to two hyperedges, $\alpha$ and $\beta$, hence it participates two games in both hyperedges. For public goods game (PGG) in hyperedge $\alpha$, vertex 2, as a cooperator, pays a cost $c$ and gets a benefit $b/g_\alpha$ from each cooperator (including vertex 2 itself) in the same hyperedge, where $g_\alpha$ is the order of hyperedge $\alpha$ and $r = b/c$ is the enhancement factor. **C,** The payoff of vertex 2 comes from the weighted average of its all neighbors. For example, vertex 3 exerts effect on vertex 2 through hyperedge $\alpha$ and $\beta$ with weights $w_\alpha/(w_\alpha + w_\beta) = 2/3$ and $w_\beta/(w_\alpha + w_\beta) = 1/3$ respectively. **D,** We construct an *interaction* graph by extracting payoff coefficients and giving each vertex the same degree as its hyperdegree. And the edge weights become the transition probabilities between vertices times the starting vertex's degree. **E,** In a random walk with self-loops, vertex 2 chooses the adjacent hyperedge $\alpha$ with probability $2/3$ or $\beta$ with probability $1/3$. **F,** Vertex 2 then traverses the members (including itself) in the hyperedge equally. **G,** The transition probability of vertex 2 is obtained by renormalizing over the weights of its adjacent hyperedges. **H**, We also give the same degree as its hyperdegree to each vertex to construct the effective graph with self-loops, $\hat{G}$, which is exactly identical with the *interaction* graph. **I-L**, The random walk with no self-loops is another akin process, with the difference that the vertex cannot return itself in one step. And the effective graph with no self-loops is obtained as $\tilde{G}$. With (**H**) and (**L**), we define an $(n, m)$-random walk to represent a path of $n$ hops on graph $\tilde{G}$ followed by $m$ hops on graph $\hat{G}$.

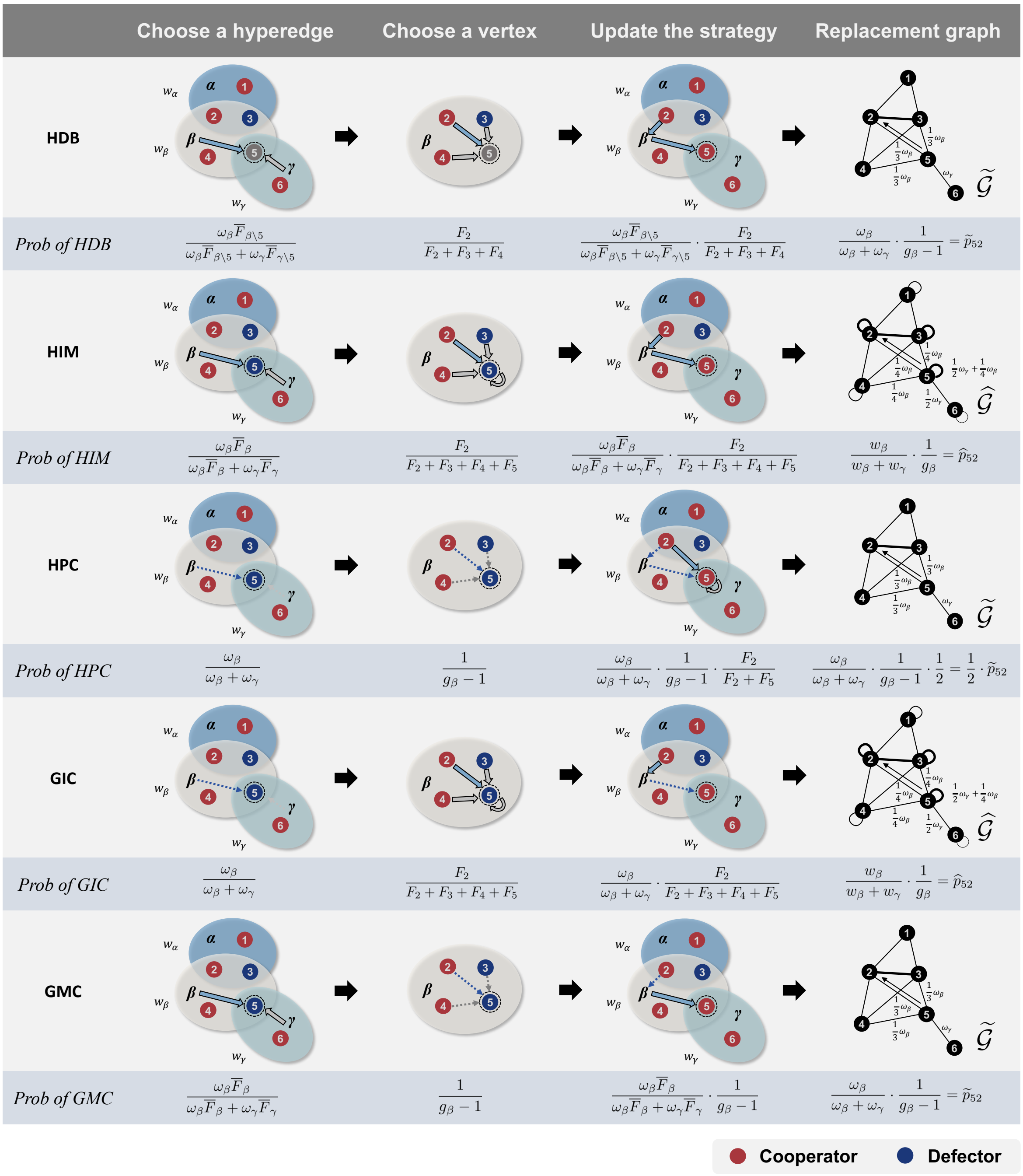

| | Choose a hyperedge | Choose a vertex | Update the strategy | Replacement graph |
|---|---|---|---|---|
| *Prob of HDB* | $\frac{\omega_\beta \overline{F}_{\beta\backslash 5}}{\omega_\beta \overline{F}_{\beta\backslash 5}+\omega_\gamma \overline{F}_{\gamma\backslash 5}}$ | $\frac{F_2}{F_2+F_3+F_4}$ | $\frac{\omega_\beta \overline{F}_{\beta\backslash 5}}{\omega_\beta \overline{F}_{\beta\backslash 5}+\omega_\gamma \overline{F}_{\gamma\backslash 5}}\cdot\frac{F_2}{F_2+F_3+F_4}$ | $\frac{\omega_\beta}{\omega_\beta+\omega_\gamma}\cdot\frac{1}{g_\beta-1}=\widetilde{p}_{52}$ |
| *Prob of HIM* | $\frac{\omega_\beta \overline{F}_\beta}{\omega_\beta \overline{F}_\beta+\omega_\gamma \overline{F}_\gamma}$ | $\frac{F_2}{F_2+F_3+F_4+F_5}$ | $\frac{\omega_\beta \overline{F}_\beta}{\omega_\beta \overline{F}_\beta+\omega_\gamma \overline{F}_\gamma}\cdot\frac{F_2}{F_2+F_3+F_4+F_5}$ | $\frac{w_\beta}{w_\beta+w_\gamma}\cdot\frac{1}{g_\beta}=\widehat{p}_{52}$ |
| *Prob of HPC* | $\frac{\omega_\beta}{\omega_\beta+\omega_\gamma}$ | $\frac{1}{g_\beta-1}$ | $\frac{\omega_\beta}{\omega_\beta+\omega_\gamma}\cdot\frac{1}{g_\beta-1}\cdot\frac{F_2}{F_2+F_5}$ | $\frac{\omega_\beta}{\omega_\beta+\omega_\gamma}\cdot\frac{1}{g_\beta-1}\cdot\frac{1}{2}=\frac{1}{2}\cdot\widetilde{p}_{52}$ |
| *Prob of GIC* | $\frac{\omega_\beta}{\omega_\beta+\omega_\gamma}$ | $\frac{F_2}{F_2+F_3+F_4+F_5}$ | $\frac{\omega_\beta}{\omega_\beta+\omega_\gamma}\cdot\frac{F_2}{F_2+F_3+F_4+F_5}$ | $\frac{w_\beta}{w_\beta+w_\gamma}\cdot\frac{1}{g_\beta}=\widehat{p}_{52}$ |
| *Prob of GMC* | $\frac{\omega_\beta \overline{F}_\beta}{\omega_\beta \overline{F}_\beta+\omega_\gamma \overline{F}_\gamma}$ | $\frac{1}{g_\beta-1}$ | $\frac{\omega_\beta \overline{F}_\beta}{\omega_\beta \overline{F}_\beta+\omega_\gamma \overline{F}_\gamma}\cdot\frac{1}{g_\beta-1}$ | $\frac{\omega_\beta}{\omega_\beta+\omega_\gamma}\cdot\frac{1}{g_\beta-1}=\widetilde{p}_{52}$ |

**Table. 1. Transition probability for five updating rules.** During the process of strategy update there are three sub-steps: choosing an adjacent hyperedge, choosing a vertex in the hyperedge to learn from, and making a decision of whether to update strategy. The transition probability of each sub-step is shown. Here $\overline{F}_{e\backslash i}$ represents the average fitness of hyperedge $e$ excluding vertex $i$ and $\overline{F}_e$ denotes the average fitness of hyperedge $e$.

**A**

| Update rules | Critical enhancement factor, $r^*$ |
|---|---|
| HDB | $\left(\tilde{N}-2\right)\Big/\left(\tilde{N}/\langle g\rangle+\tilde{N}\theta^{(1,1)}-2\right)$ |
| HIM | $\left(\tilde{N}+\tilde{N}/\langle g\rangle-2\right)\Big/\left(\tilde{N}/\langle g\rangle+\tilde{N}\theta^{(0,2)}-2\right)$ |
| HPC | $\left(\tilde{N}-1\right)\Big/\left(\tilde{N}/\langle g\rangle-1\right)$ |
| GIC | $\left(\tilde{N}-1\right)\Big/\left(\tilde{N}/\langle g\rangle-1\right)$ |
| GMC | $\left(\tilde{N}/\langle g\rangle-1\right)\Big/\left(\tilde{N}\theta^{(0,2)}-1\right)$ |

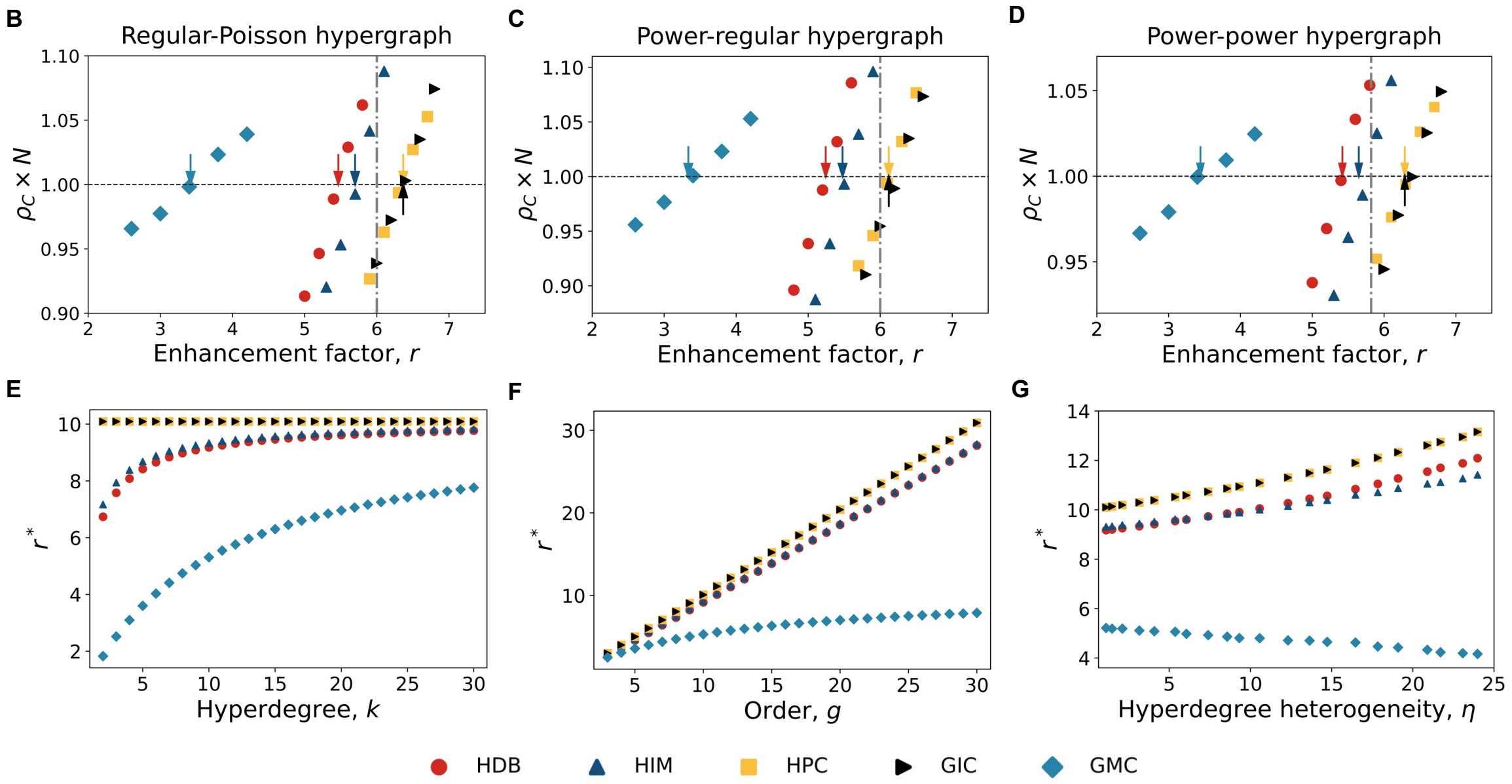

**Fig. 2. Critical enhancement factors for cooperation emergence in hypergraphs. a**, The critical thresholds $r^*$ for the five updating rules. Here $\langle x\rangle$ and $\langle x^2\rangle$ represent the first and second moments of variable $x$ respectively; $\tilde{N}=N/\eta$ where $\eta=\langle k^2\rangle/\langle k\rangle^2$ indicates the hyperdegree heterogeneity and $N$ is the number of vertices in the system; $\theta^{(n,m)}$ represents the hyperdegree-weighted probability of random walks from any vertex to itself along paths involving $n$ hops in effective graph $\tilde{G}$ (with no self-loops) and $m$ hops on effective graph $\hat{G}$ (with self-loops). **b-d**, Numerical validations for the theoretical results of critical enhancement factors. The size of the three hypergraphs is $N=100$, and the mean hyperdegree and mean order are around 6. We consider different distributions of hyperedges' orders and vertices' hyperdegrees. For example, regular-Poisson hypergraph (**b**) means that the hyperedges have the same order and the vertices' hyperdegrees obey a Poisson distribution. Horizontal dashed lines represent the fixation probability baseline, and vertical dashed lines represent the mean order of hyperedges $\langle g\rangle$. Each dot represents the fixation probability times population size out of $5\times10^6$ independent simulations under weak selection $\delta=0.025$. The arrows mark the theoretical results of critical enhancement factors. **e**, Regular hypergraphs (all vertices have the same hyperdegree and all hyperedges have the same order) with 1000 vertices. The hyperedge order is fixed to 10 and the vertex hyperdegree varies. **f**, Similar to (**e**) but the vertex hyperdegree is fixed to 10 and the hyperedge order varies. **g**, The averages of the order and hyperdegree are around 10 yet the hyperdegrees obey different distributions.

| Hypernetwork | Weighted? | $N$ | $L$ | $\langle k \rangle$ | $\langle g \rangle$ | Rule | $r^*$ | $\rho_C(r^*) \times N$ |
|---|---|---|---|---|---|---|---|---|
| Coauth-DBLP | yes | 207 | 61 | 1.3140 | 4.4590 | HDB | 2.6273 | 0.9963 |
| | | | | | | HIM | 3.0969 | 0.9912 |
| | | | | | | HPC | 4.5599 | 0.9956 |
| | | | | | | GIC | 4.5599 | 1.0012 |
| | | | | | | GMC | 1.2574 | 1.0009 |
| Coauth-MAG | yes | 221 | 52 | 1.2670 | 5.3846 | HDB | 3.1494 | 0.9983 |
| | | | | | | HIM | 3.6270 | 0.9875 |
| | | | | | | HPC | 5.5328 | 1.0088 |
| | | | | | | GIC | 5.5328 | 1.0016 |
| | | | | | | GMC | 1.2482 | 0.9970 |
| Threads-math | no | 118 | 131 | 2.2288 | 2.0076 | HDB | 1.4350 | 1.0027 |
| | | | | | | HIM | 1.8231 | 0.9957 |
| | | | | | | HPC | 2.0969 | 0.9956 |
| | | | | | | GIC | 2.0969 | 0.9921 |
| | | | | | | GMC | 1.4311 | 0.9933 |
| Threads-ubuntu | no | 117 | 117 | 2.0171 | 2.0171 | HDB | 1.3858 | 0.9911 |
| | | | | | | HIM | 1.7912 | 1.0004 |
| | | | | | | HPC | 2.0929 | 0.9807 |
| | | | | | | GIC | 2.0929 | 0.9995 |
| | | | | | | GMC | 1.3760 | 1.0121 |

**Table. 2. Empirical hypergraphs.** The properties of four empirical hypergraphs. Here the values of $r^*$ are our mathematical results for the critical enhancement factors, and $\rho_C(r^*) \times N$ denotes the simulation results. Here, $\rho_C(r^*) \times N = 1.0$ indicates a perfect agreement between our theoretical results and numerical simulations.

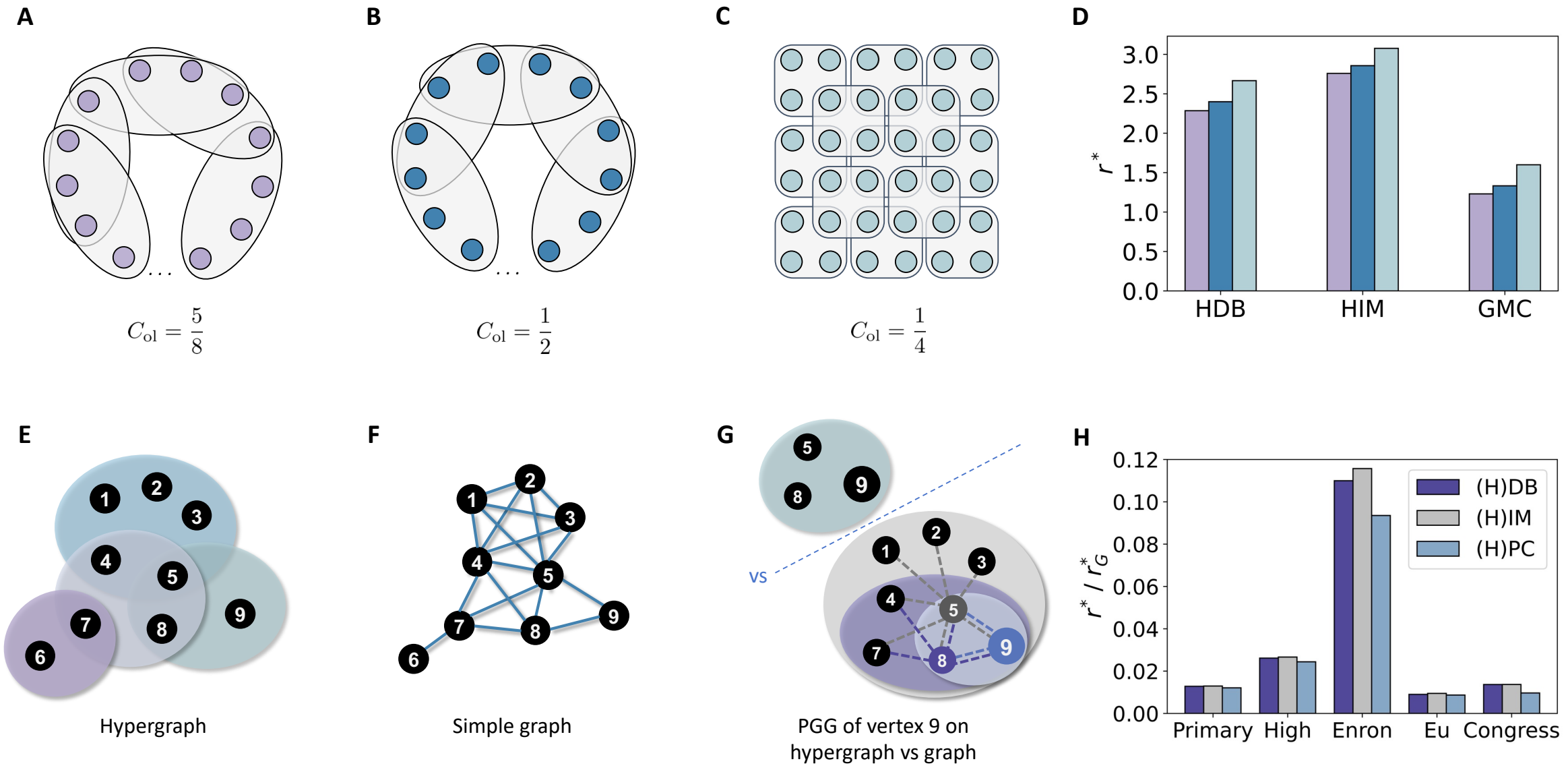

**Figure 3 | Impact of hyperedge overlaps and comparison between pairwise and higher-order interactions. a-c,** For the three regular hypergraphs, the hyperedge overlap are $5/8$, $1/2$, and $1/4$ respectively, according to the definition described in Methods. **d,** Critical enhancement factors for different strengths of overlap, under HDB, HIM and GMC updating rules. **e,** A hypergraph with 9 vertices. **f,** The *trivially* mapped simple graph obtained by just retaining the vertex-to-vertex connections. **g,** For PGG on hypergraph, vertex 9 involves in the game occurring in the hyperedge it belongs to. In contrast, for PGG on simple graph[14], vertex 9 participates the games of three vertex sets, i.e., vertex 9 and its neighbors, vertex 5 and its neighbors, 8 and its neighbors. **h,** The ratio between the critical enhancement factors $r^*$ of four empirical hypergraphs and the ciritical enhancement factors $r_G^*$ of their *trivially* mapped simple graphs.

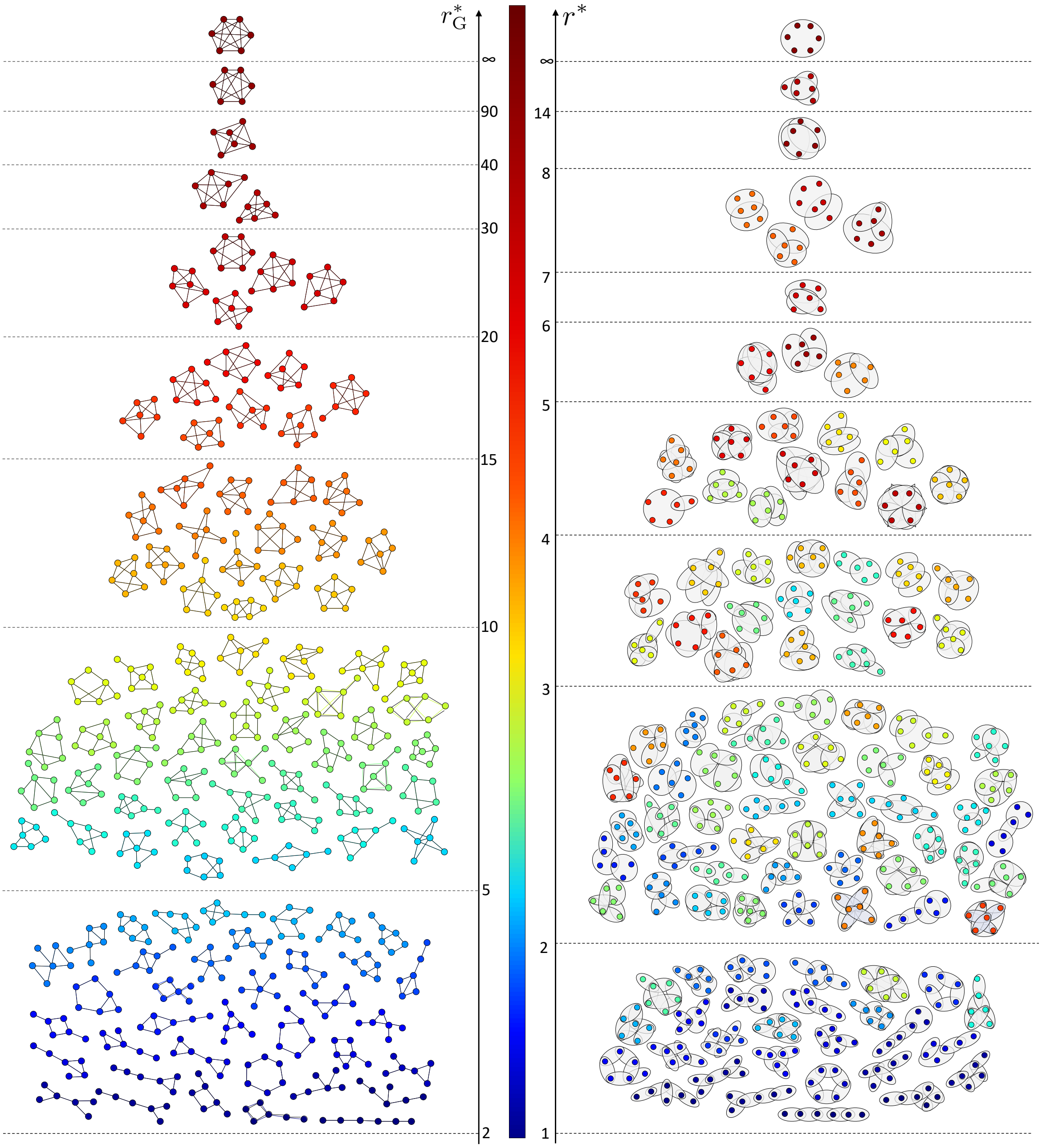

**Fig. 4. Comparison between simple graphs and corresponding hypergraphs for updating rule DB.** From each simple graph the cliques are extracted to construct hyperedges, forming the corresponding hypergraph. A simple graph and its corresponding hypergraph take the same RGB value. The simple graphs and hypergraphs are sorted according to their critical enhancement factors.